\documentclass[twocolumn,showpacs,aps,prl]{revtex4}

\usepackage[english]{babel}
\usepackage{amsfonts, amsmath, amstext, amssymb, amsfonts, amsxtra}
\usepackage{braket}
\usepackage{bbm}
\usepackage{bbold}
\usepackage{wasysym}
\usepackage{graphicx}

\DeclareMathOperator{\tr}{tr} 
\newcommand{\im}{{i}}         
\newcommand{\St}{\mathcal{S}} 

\begin{document}
\title{Quantum bifurcation diagrams}
\author{M.~Ivanchenko$^1$, E.~Kozinov$^2$, V.~Volokitin$^3$, A.~Liniov$^{2,3}$, I.~Meyerov$^{2,3}$, and S.~Denisov$^{1,4}$}
\affiliation{
$^{1}$Department of Applied Mathematics, Lobachevsky State University of Nizhny Novgorod, Gagarina Av.\ 23, Nizhny Novgorod, 603950, Russia
$^{2}$Department of Mathematical Software and Supercomputing Technologies, Lobachevsky University, Gagarina Av.\ 23, Nizhny Novgorod, 603950, Russia\\
$^{3}$Institute of Supercomputing Technologies, Lobachevsky University, Gagarina Av.\ 23, Nizhny Novgorod, 603950, Russia \\
$^{4}$Institute of Physics, University of Augsburg, Universit\"{a}tsstra{\ss}e 1, 86159 Augsburg, Germany \\
}

\begin{abstract}

 Asymptotic state of an open quantum system can undergo qualitative changes upon small variation of system parameters. We demonstrate it that such 'quantum bifurcations' can be appropriately defined and made visible as changes in the structure of the asymptotic density matrix. By using an $N$-boson open quantum dimer, 
 we present quantum diagrams for the pitchfork and saddle-node bifurcations in the stationary case and visualize a period-doubling transition to chaos for the periodically modulated dimer. In the latter case, we also identify a specific bifurcation of purely quantum nature. 
	
\end{abstract}
\maketitle

\section{Introduction}

Correspondence between quantum systems and their classical counterparts (including different mean-filed approximations) is the core issue in the quantum chaos theory \cite{stock}. There these relations are mainly analyzed in terms of spectral characteristics of quantum Hamiltonians; one of the milestones was linking the spectral statistics and chaos -- regular dynamics transition in the phase space of the corresponding classical system \cite{stock}.  Quantum footprints of bifurcations, sudden qualitative changes in the phase-space structure of a classical Hamiltonian system upon a small variation of parameter(s), were also intensively explored (through relatively recently) \cite{bif0,bif1,bif2}.  
It was found that such archetypical bifurcations as the pitchfork  and Hopf bifurcations \cite{bifC} in  classical Hamiltonian mean-field equations are connected to sharp changes of the ground-state entanglement in the corresponding quantum models \cite{bif0,bif2}. A pitchfork bifurcation was also found to be responsible for the transition from Rabi to Josephson dynamics in experiments with rubidium spinor Bose-Einstein condensate \cite{bif3}.

In the context of open quantum systems, a further extension of  quantum-classical bifurcation connections is challenging. First, bifurcations could make a stronger impact there than in the Hamiltonian case because
they will affect the stationary state of the system as a whole, not only a specific eigenstate. Second, it will be an important step towards the foundation of \textit{disspative} quantum chaos, a theory,
which is still at the  beginning of its development. However, open (or 'dissipative')  quantum systems \cite{book} are much less explored in this respect, primarily, due to the difficulties in constructing proper mean-filed equations for open quantum models. Even when the description is restricted to the established Markovian framework
and the dynamics of a model  is described by the Lindblad equation \cite{lind,gorini, alicki,book}, it is not so apparent how to recast the system dynamics in mean-filed terms.
Usually it is done in the spirit of Bogoliubov-Born-Green-Kirkwood-Yvon (BBGKY) approach by truncating the hierarchy of cumulants right on the level of expectation values \cite{bg3} or keeping double correlators \cite{bg1,bg2}. This way, a quantum variant of the pitchfork bifurcation was found following in Ref.~\cite{bg3}, where it was also visualized by means of the Wigner distribution.

Here we propose an upside-down approach. Namely, our idea is that bifurcations can be seen in the original quantum system, and then, in case of a need, 
can be compared with bifurcations in a constructed mean-filed model. That is, for a particular many-body model that we consider, an open dimer 
with $N$ interacting bosons \cite{coherent, Vardi, Witthaut, PolettiKollath2012, map}, a bifurcation diagram can be built plotting diagonal elements of the asymptotic density matrix. We compare the 'quantum bifurcation diagrams' constructed in this way with the diagrams obtained from mean-field equations and report both similarities and differences between the two.

\section{Model}

We consider a system of $N$ interacting bosons hopping over a dimer that is periodically
driven, described by a time-periodic Hamiltonian
\begin{equation}
H(t)=J \left( b_1^{\dagger} b_2 + b_2^{\dagger} b_1 \right) + \frac{2U}{N} \sum_{j=1,2} n_j\left(n_j-1\right)+\varepsilon(t)\left(n_2 - n_1\right) 
\label{eq:2}
\end{equation}
where $J$ is the tunneling amplitude, $U/N$ the interaction strength normalized by a number of particles, and
$\varepsilon(t)$ presents the modulation of the local potential. We
choose $\varepsilon(t)=\varepsilon(t+T)=E+A\theta(t)$, where
$E$ and $A$ are a static and a dynamic energy offset
between the two sites, respectively, and the piecewise constant periodic driving, $\theta(t)=1$ for $0\le t<T/2$, $\theta(t)=0$ for $T/2\le t<T$. Here, $b_j$ and $b_j^{\dagger}$ are the annihilation and
creation operators of an atom at site $j$, while $n_j=b_j^{\dagger}b_j$. This
Hamiltonian has been studied theoretically \cite{coherent, Vardi, Witthaut,
PolettiKollath2012} and was implemented in several experimental studies
\cite{oberthaler, ober1}. 

The system is open, so its dynamics is  modeled with a quantum master equation whose generator
$\mathcal{L}$ is of Lindblad form \cite{lind,gorini, alicki,book}
\begin{equation}
\label{eq:1}
\dot{\varrho} = \mathcal{L}_t(\varrho) = -\im [H(t),\varrho] + \mathcal{D}(\varrho).
\end{equation}


For a dissipator a single jump operator is used \cite{DiehlZoller2008, zoller},
\begin{equation}
\begin{aligned}
&\mathcal{D}(\varrho) = \frac{\gamma}{N} \left[V\varrho V^\dagger - \frac{1}{2}\{V^\dagger V,\varrho\}\right],\\
&V=(b_1^{\dagger} + b_2^{\dagger})(b_1-b_2).
\end{aligned} 
\label{eq:3}
\end{equation}
Its action tries to `synchronize' the dynamics on the sites by constantly recycling
anti-symmetric out-phase modes into the symmetric in-phase ones. The scaled coupling constant
$\gamma/N$ is assumed to be time-independent. Weak dissipation limit will be  addressed so we set $\gamma=0.1$.
As a Fock basis we will take the states corresponding to a certain number of bosons, $i$, on the first site, $i = 0,...,N$.

To relate the quantum and classical bifurcations,  a set
of mean-field equations can be derived \cite{map} and then its attractor solutions can be compatred  with the solutions of the
original quantum system. For the dimer problem, one rewrites the master equation (\ref{eq:1}) in terms of the spin
operators

\begin{figure*}[th!!!]
(a)  \includegraphics[angle=270, width=0.95\columnwidth]{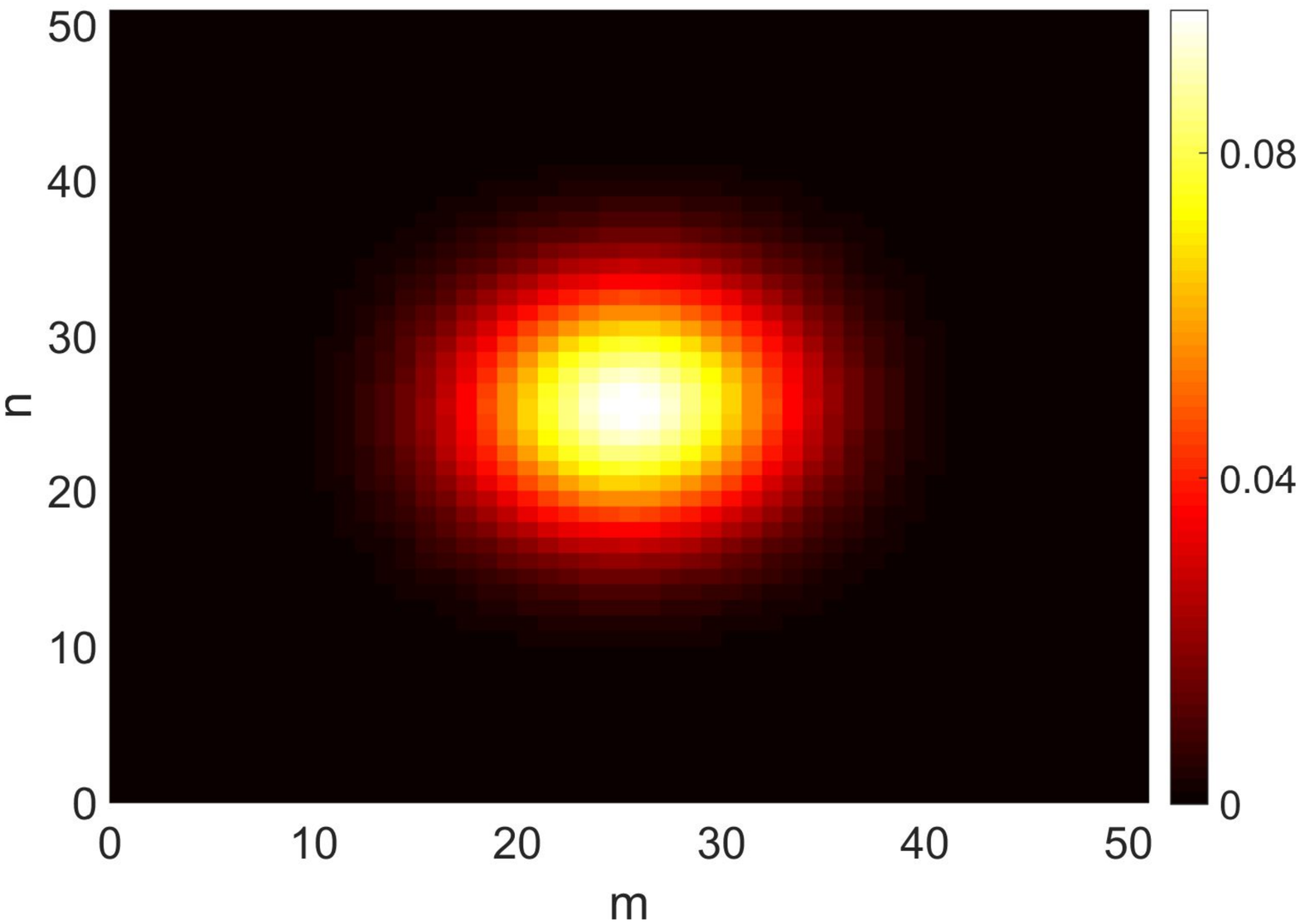}%
(b)  \includegraphics[angle=270,width=0.95\columnwidth]{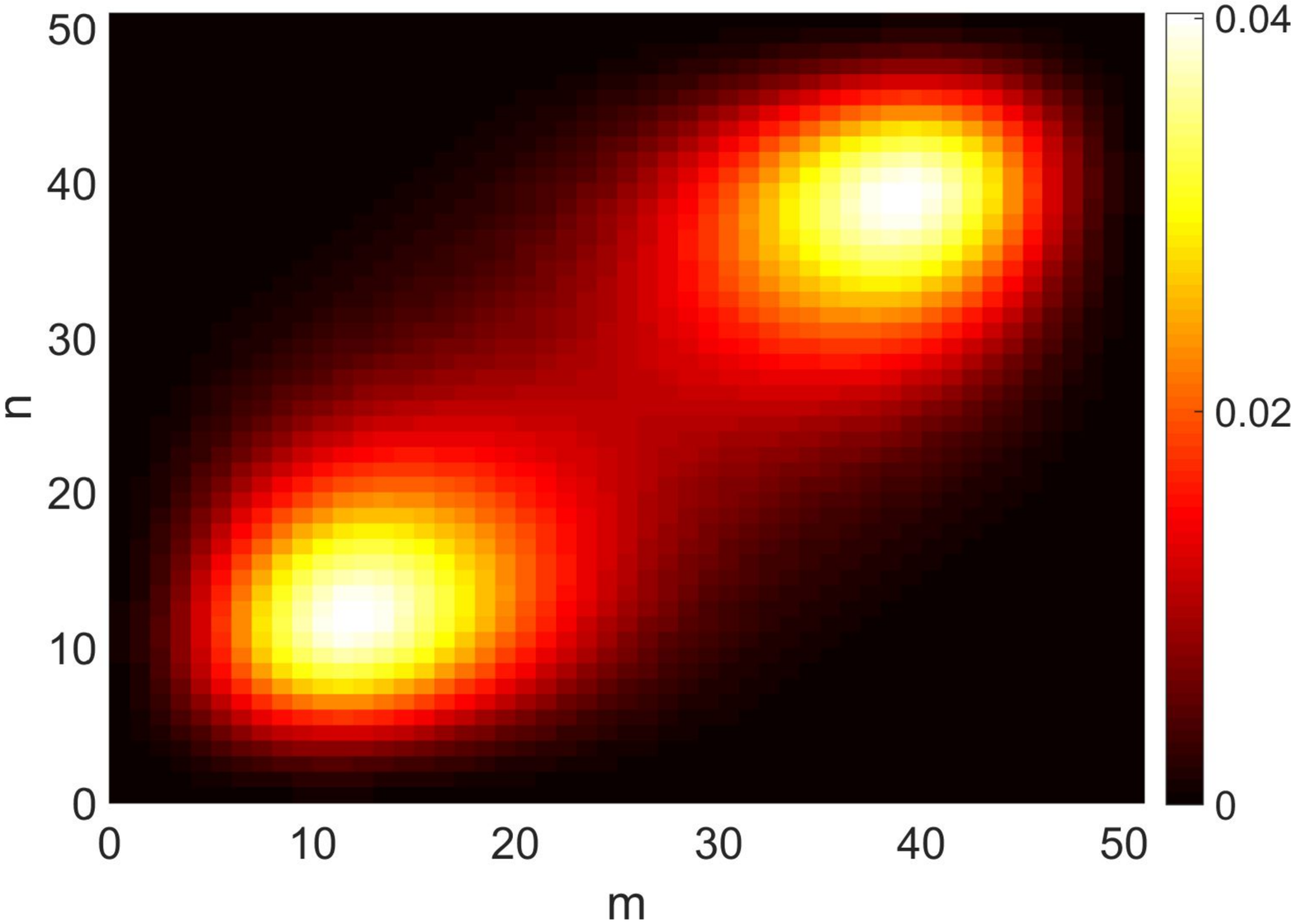}%
	\caption{\label{fig:1}
    Quantum pitchfork bifurcation: as interaction $U$ increases, 
    the distribution of absolute values of the elements,  $|{\varrho}_{m,n}|$, of the stationary density matrix, 
    evolves from a unimodal, $U=0.2$ (a) to a bimodal, $U=0.6$ (b). The  parameters are  $J=1, E=0, \gamma=0.1, N=50$.}
\end{figure*}

$\St_x=\frac{1}{2N}\left(b^{\dagger}_1 b_2 + b^{\dagger}_2 b_1\right)$,
$\St_y= -\frac{\im}{2N}\left(b^{\dagger}_1 b_2 - b^{\dagger}_2 b_1\right)$,
$\St_z=\frac{1}{2N}\left(n_1 - n_2\right)$, and then considers their evolution in the
Heisenberg picture \cite{book}. For a large number of atoms $N$, the commutator
$\left[\St_x,\St_y\right]=\im {\St_z}/{N} {\overset{N\to\infty}{=}} 0$
and similarly for other cyclic permutations. Replacing operators with their
expectation values, $\braket{\St_k} =\tr [\varrho \St_k]$, and denoting
$\braket{\St_k}$ by $S_k$, we arrive at \cite{map} 
\begin{equation}
\begin{aligned}
&\frac{\mathrm{d} S_x}{\mathrm{d}t} = 2\varepsilon(t)S_y - 8U S_zS_y + 8\gamma \left(S_y^2+S_z^2\right),\\
&\frac{\mathrm{d} S_y}{\mathrm{d}t} = -2\varepsilon(t)S_x + 8U S_xS_z -2JS_z - 8\gamma  S_xS_y.\\
&\frac{\mathrm{d} S_z}{\mathrm{d}t} = -2JS_y - 8\gamma S_xS_z,
\end{aligned}
\label{eq:4}
\end{equation}
As the quantity $S^2=S_x^2+S_y^2+S_z^2$ is a constant of motion, 
the mean-field evolution can be reduced to the surface of a Bloch sphere,
$\left(S_x,S_y,S_z\right) = \frac{1}{2} [\cos(\varphi)\sin(\vartheta),\sin(\varphi)\sin(\vartheta), \cos(\vartheta) ]$,
yielding the equations of motion \cite{map}
\begin{equation}
\begin{aligned}
&\dot{\vartheta} = -2J\sin(\varphi) + 4\gamma \cos(\varphi)\cos(\vartheta), \\
&\dot{\varphi} = -2J\frac{\cos(\vartheta)}{\sin(\vartheta)}\cos(\varphi) - 2\varepsilon(t) +4U \cos(\vartheta) - 4\gamma  \frac{\sin(\varphi)}{\sin(\vartheta)}. 
\end{aligned}
\label{eq:5}
\end{equation}
The corresponding particle number on the first site is then recovered as $n=(1+\cos(\vartheta))N/2$ (index $1$ is omitted for brevity).

\begin{figure*}[th!!!]
(a)   \includegraphics[angle=270,width=0.6\columnwidth]{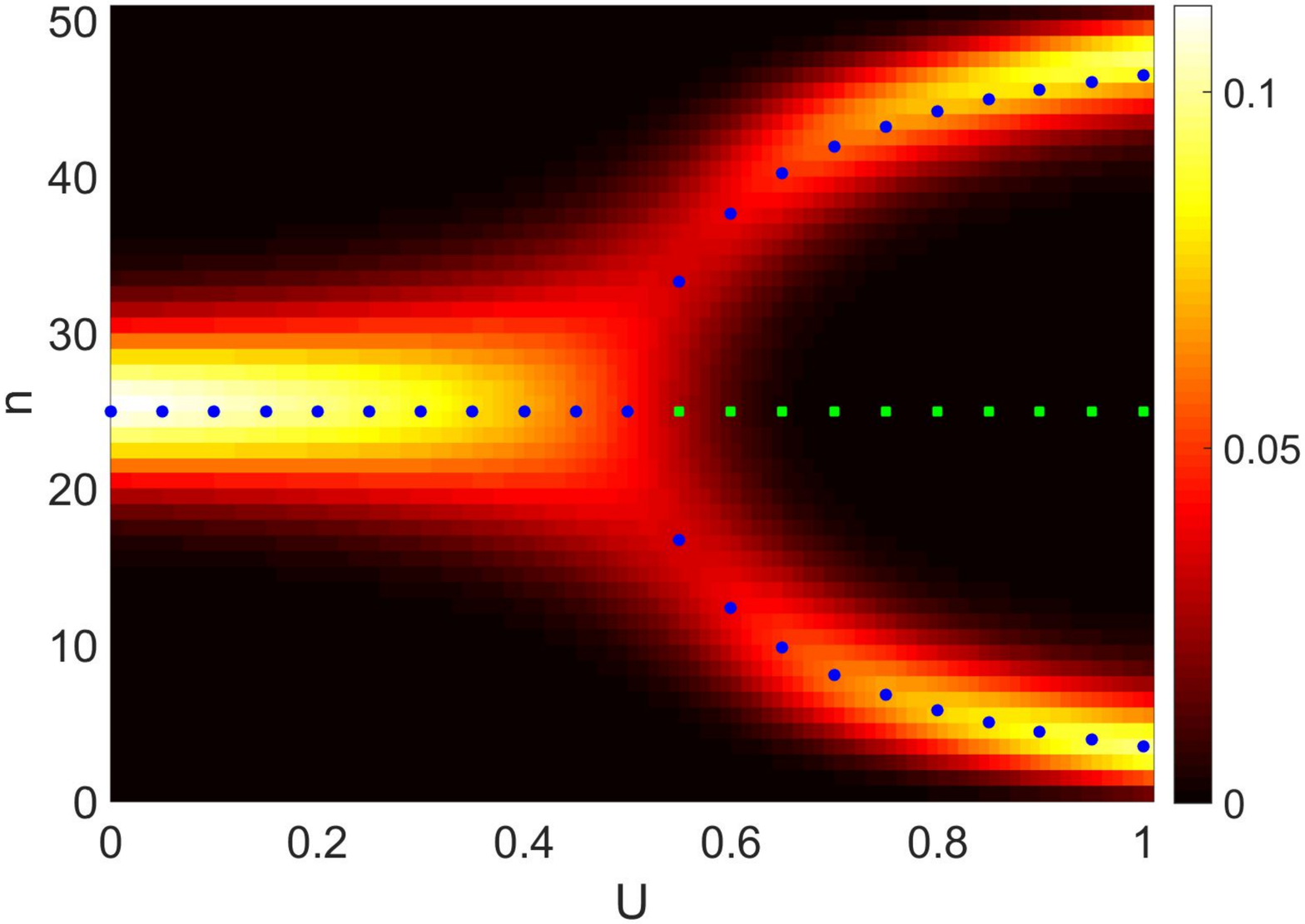}%
(b)    \includegraphics[angle=270,width=0.6\columnwidth]{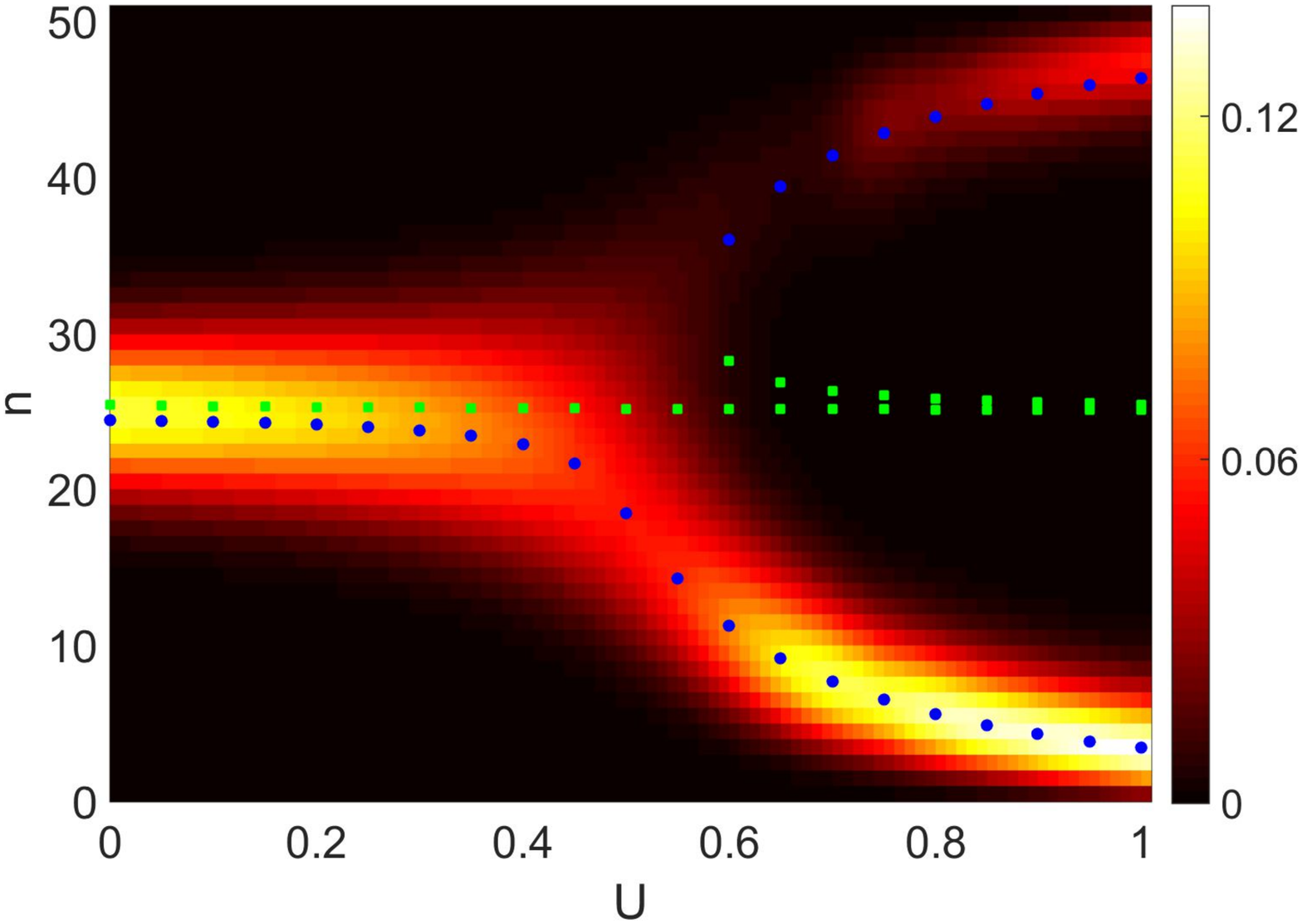}%
(c)  \includegraphics[angle=270,width=0.6\columnwidth]{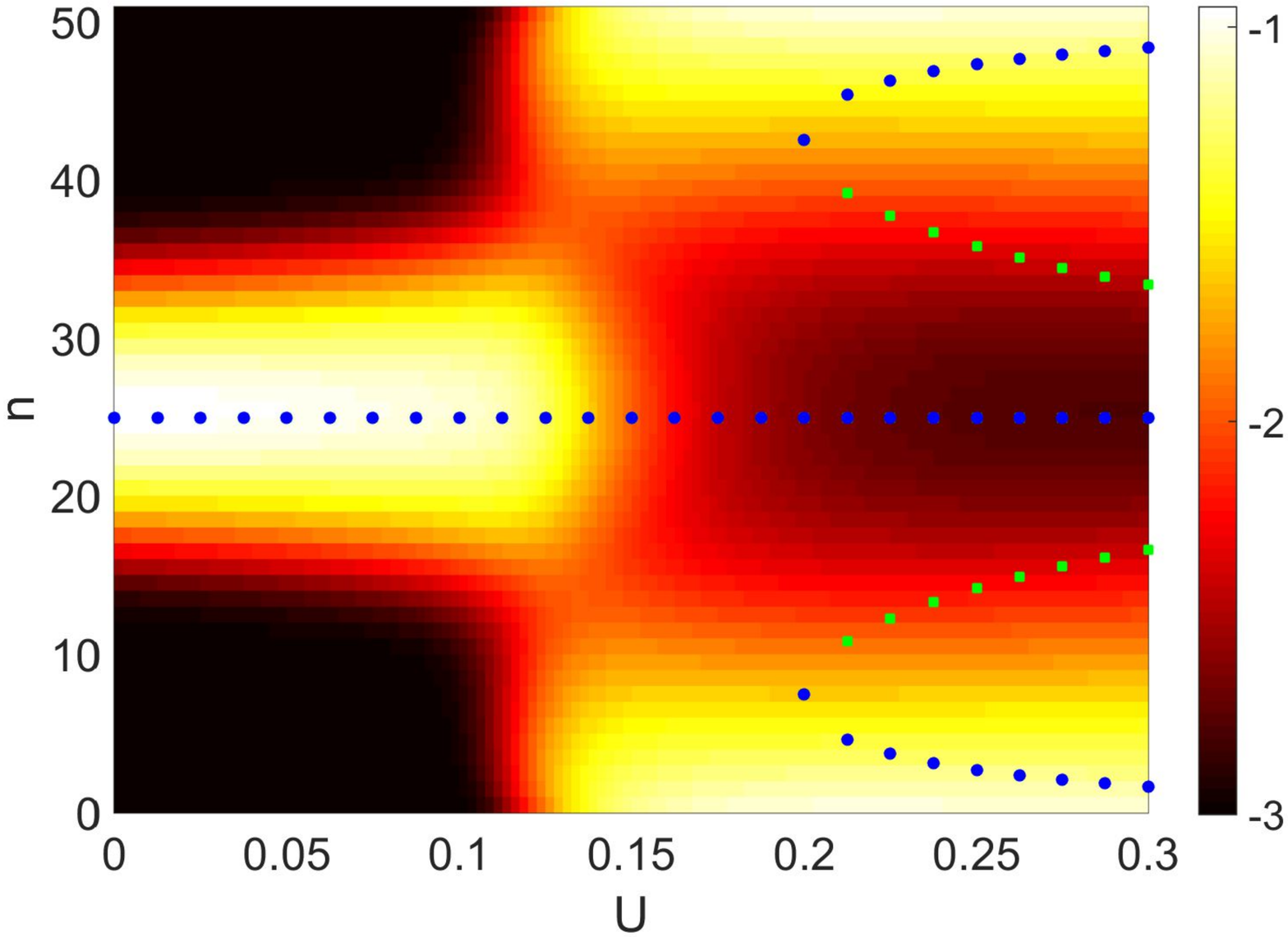}\\%
(d)  \includegraphics[angle=270,width=0.6\columnwidth]{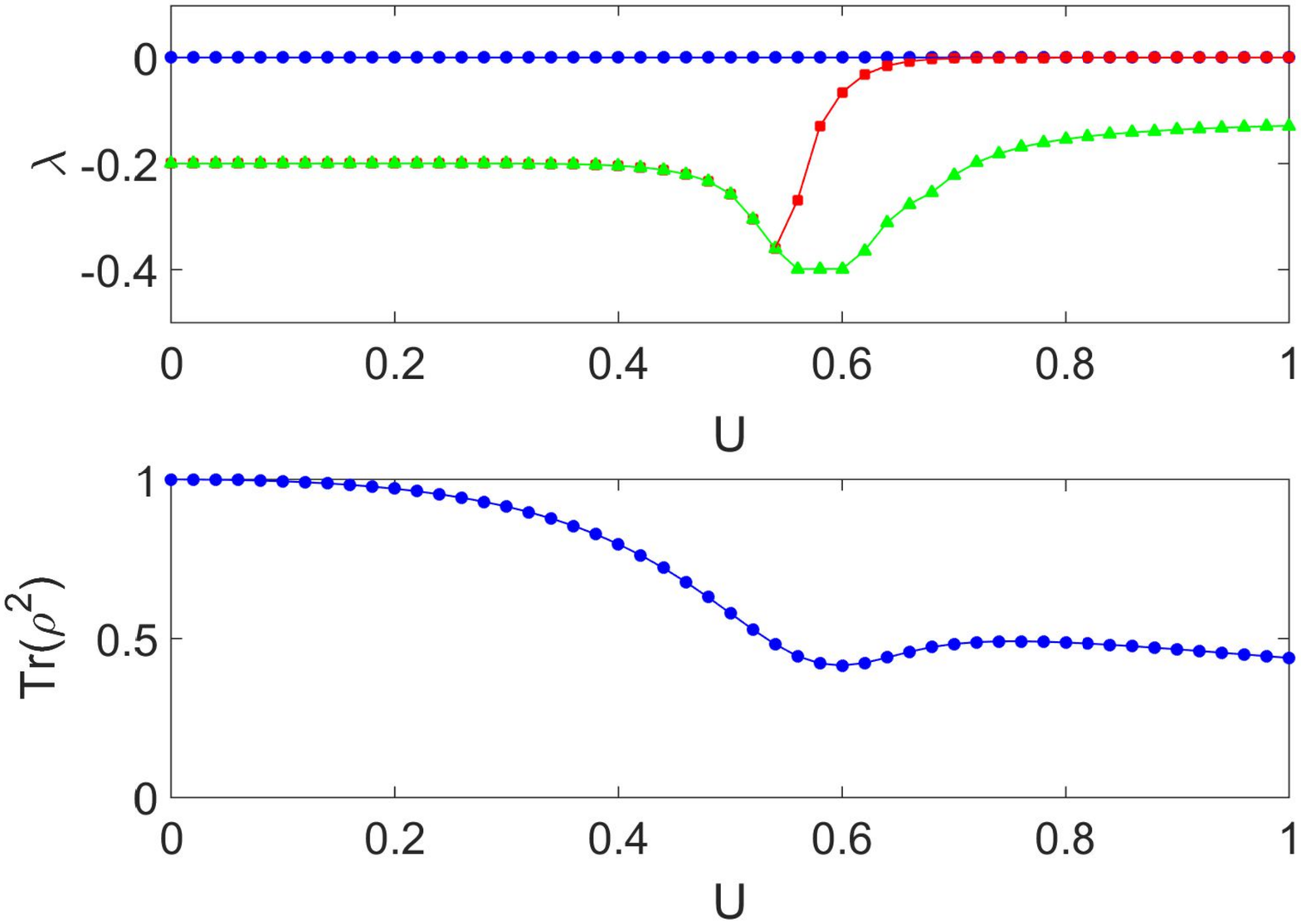}%
(e)  \includegraphics[angle=270,width=0.6\columnwidth]{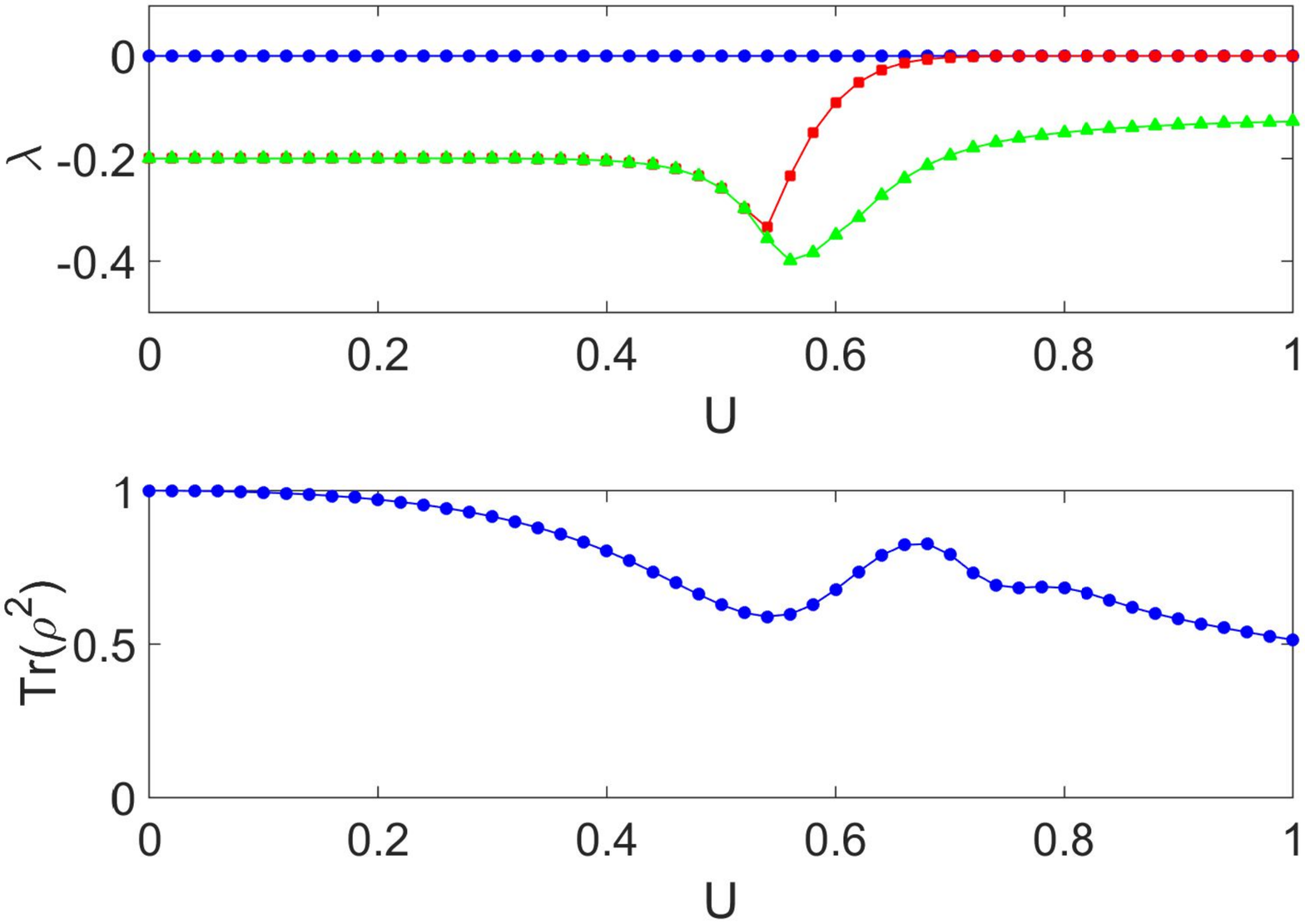}%
(f)  \includegraphics[angle=270,width=0.6\columnwidth]{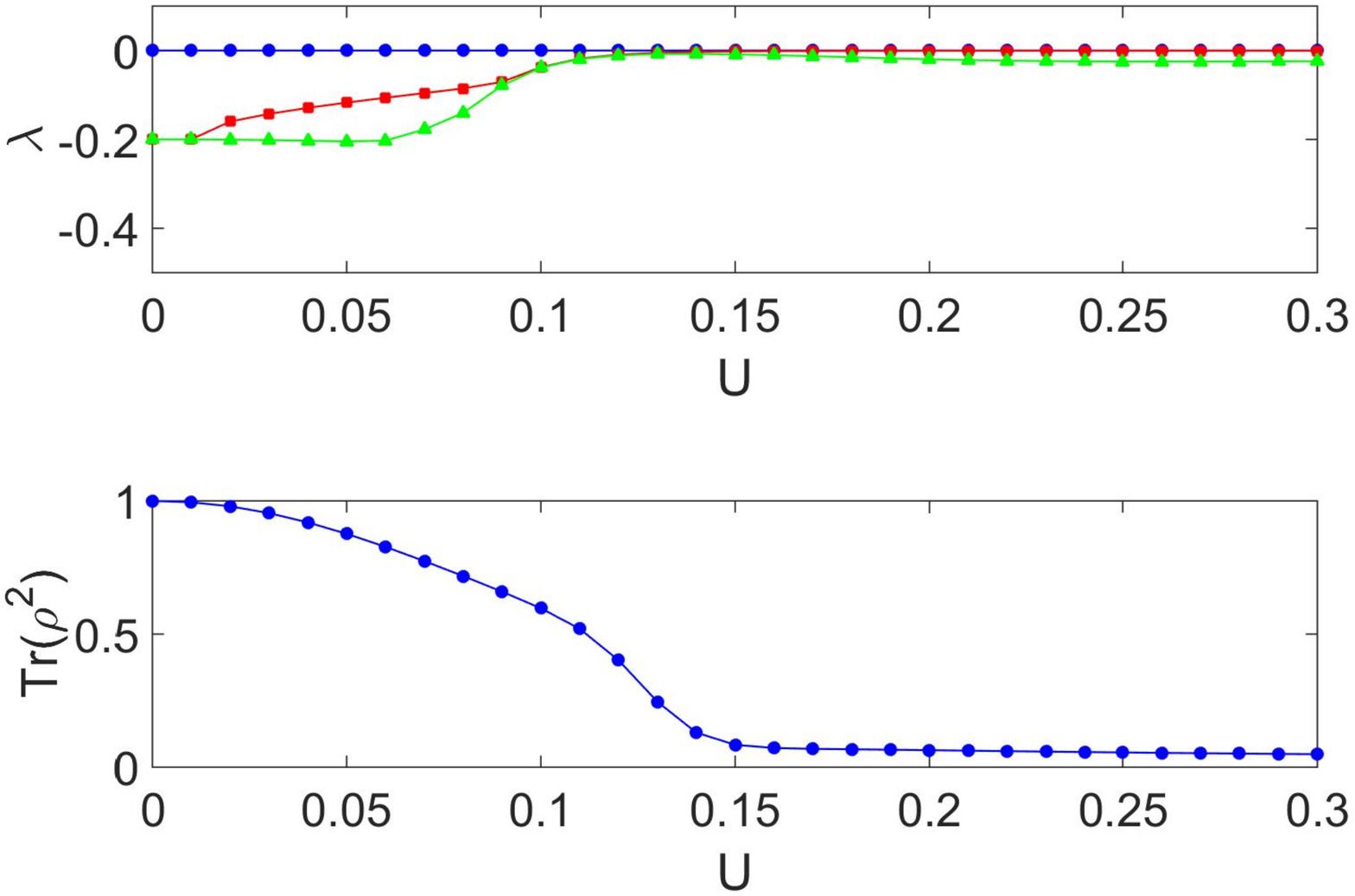}%
	\caption{\label{fig:2} 
    One-parameter bifurcation diagrams of quantum and classical models: (a) pitchfork bifurcation, $J=1, E=0$;
    (b) saddle-node bifurcation, $J=1, E=0.02$; (c) 
    a couple of simultaneous saddle-node bifurcations followed by a sub-critical pitchfork bifurcation to a 
    central state, $J=0.02, E=0$. Color encodes the diagonal elements  ${\varrho}_{n,n}$  of the stationary density matrix in (a,b) and $\log_{10}{\varrho}_{n,n}$ in (c); blue circles and green squares mark stable and unstable equilibria of the mean-field equations. (d,e,f) Corresponding three largest real parts of eigenvalues of the super-operator and purity of the stationary solution. The parameters are  $\gamma=0.1, N=50$.}
\end{figure*}

\section{Stationary bifurcations}

We start with the stationary case, $A=0$. Then, Eq.(\ref{eq:1}) can be recast in the super-operator -- super-matrix form, such that 
\begin{equation}
\label{eq:6}
\dot\varrho_s=\Pi\varrho_s,
\end{equation}
where $(\varrho_s)_{m N+n}=\varrho_{m,n}$ the indexes running $n,m=1\ldots N+1$, constitutes a super-vector, and $\Pi$ is a suitably constructed $(N+1)^2\times (N+1)^2$ constant super-matrix. 

Under general conditions, the dissipative linear system (\ref{eq:6}) has an equilibrium, $\Pi\varrho_s=0$, which is unique \cite{alicki}, 
provided normalization of the corresponding stationary density matrix, $\tr[\varrho]=1$. The spectrum of the 
super-operator, $\{\lambda_k\}$, $\Pi\varrho_s=\lambda\varrho_s$, is given by the zero 
largest eigenvalue $\lambda_1=0$, and the rest with negative real parts,  $0>{\rm Re}\lambda_2\ge\ldots{\rm Re}\lambda_{(N+1)^2}$. 
Therefore, an equilibrium state can be found as an eigenstate of the super-operator corresponding  to 
zero eigenvalue. The rest of the spectrum determines the dynamics on a way to the equilibrium.    

From the dynamical viewpoint, evolution of Eqs.~(\ref{eq:1}) and (\ref{eq:6}) is quite simple. 
A more detailed analysis, however, elucidates qualitative changes in the structure of stationary density matrix, which can be taken as signatures of quantum bifurcations. In Fig.\ref{fig:1} we plot the magnitudes of the elements of the stationary density matrix for two different values 
of particle interaction strength $U$, zero bias, $E=0$,  hopping $J=1$, and number of particles $N=50$.  
As $U$ increases, the unimodal distribution ($U=0.2$) undergoes symmetry breaking and bimodal distribution emerges ($U=0.6)$. Recalling that the element numbering in the density matrix corresponds to the particle number on the first site, we interpret this quantum bifurcation as the transition from the even splitting of particles between the sites to accumulation in one or another site.   

Following the variation of the diagonal elements, $\varrho_{n,n}$, as a function of interaction 
strength, we obtain a one-parameter quantum bifurcation diagram,  Fig.\ref{fig:2}. Moreover, calculating  the bifurcation 
diagram for the classical mean-field system, Eq.(\ref{eq:5}), we find an excellent correspondence to the supercritical 
pitchfork bifurcation, the local maxima along the diagonal of the density matrix are matching stable equilibria of the classical system, Fig.\ref{fig:2}(a).
In parallel,  the largest non-zero real part eigenvalue of the super-matrix, ${\rm Re}\lambda_2$, rapidly approaches to zero after bifurcation and then remains weakly negative as $U$ increases further, see Fig.\ref{fig:2}(d), upper panel. The bimodal distribution produced by quantum 
bifurcation is, therefore, characterized by a significantly slower relaxation rate, as well as by a lower purity, Fig.\ref{fig:2}(d), bottom panel.

\begin{figure*}[th!!!]
(a)  \includegraphics[angle=270,width=0.95\columnwidth]{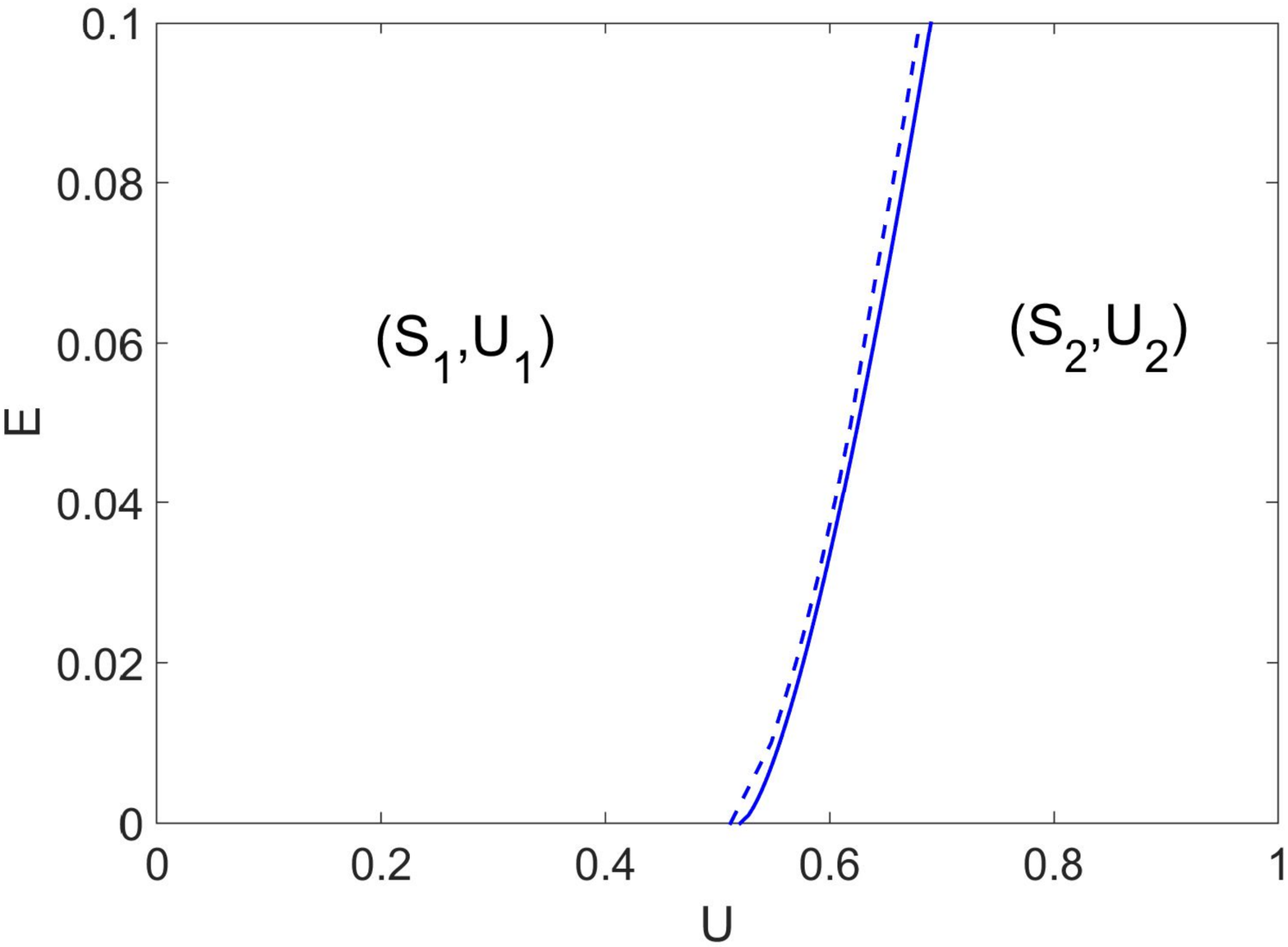}%
(b)  \includegraphics[angle=270,width=0.95\columnwidth]{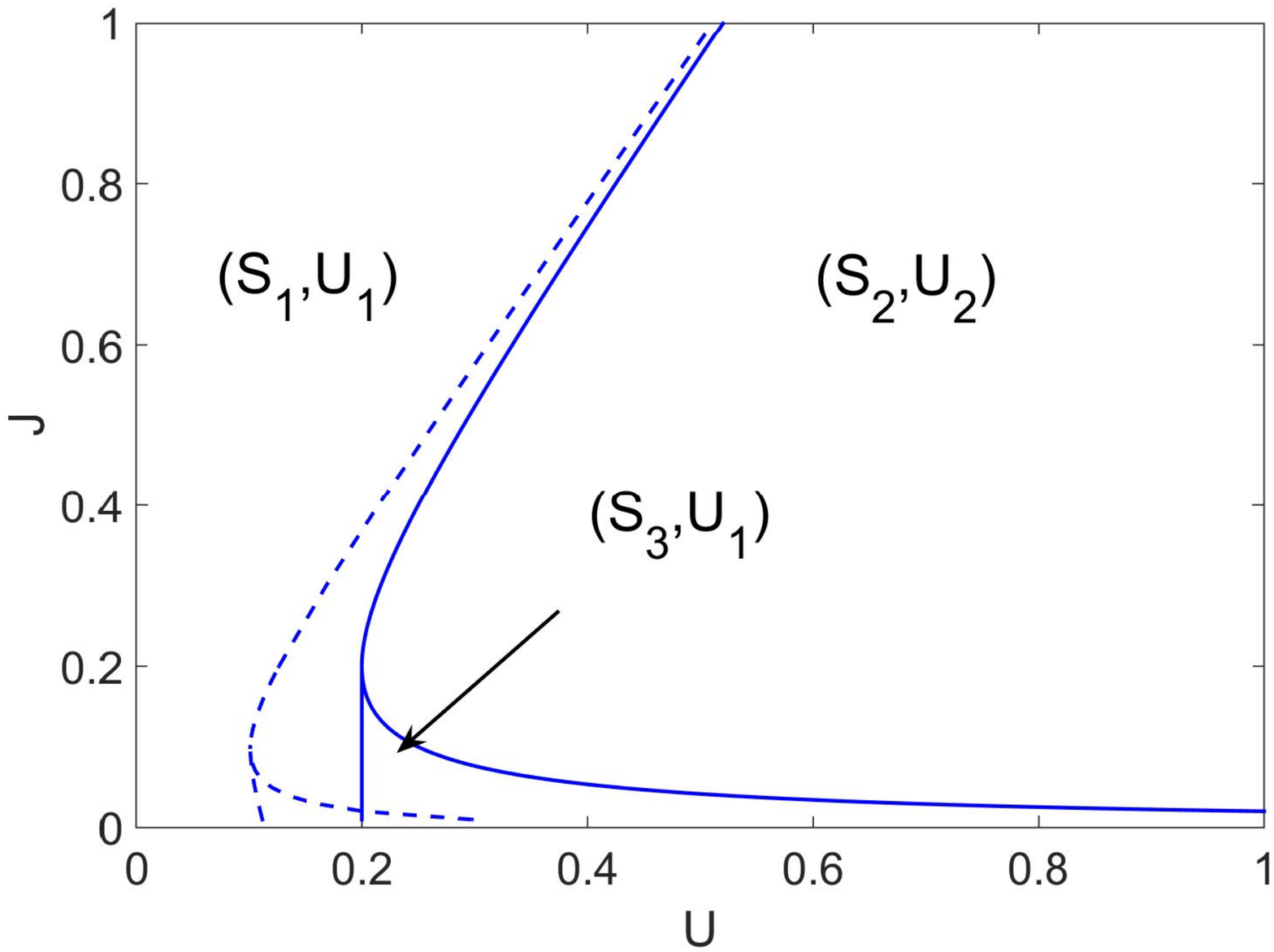}%
	\caption{\label{fig:3} 
    Two-parameter bifurcation diagrams for the quantum (dashed lines) and classical (solid lines) models, 
    (a) $J=1$ and (b) $E=0$. ${\rm (S_q,U_p)}$ denotes a region with $q$ stable and $p$ unstable equilibria in the mean-field equations. Other parameters are  $\gamma=0.1, N=50$.}
\end{figure*}

\begin{figure*}[th!!!]
(a)  \includegraphics[angle=270,width=0.95\columnwidth]{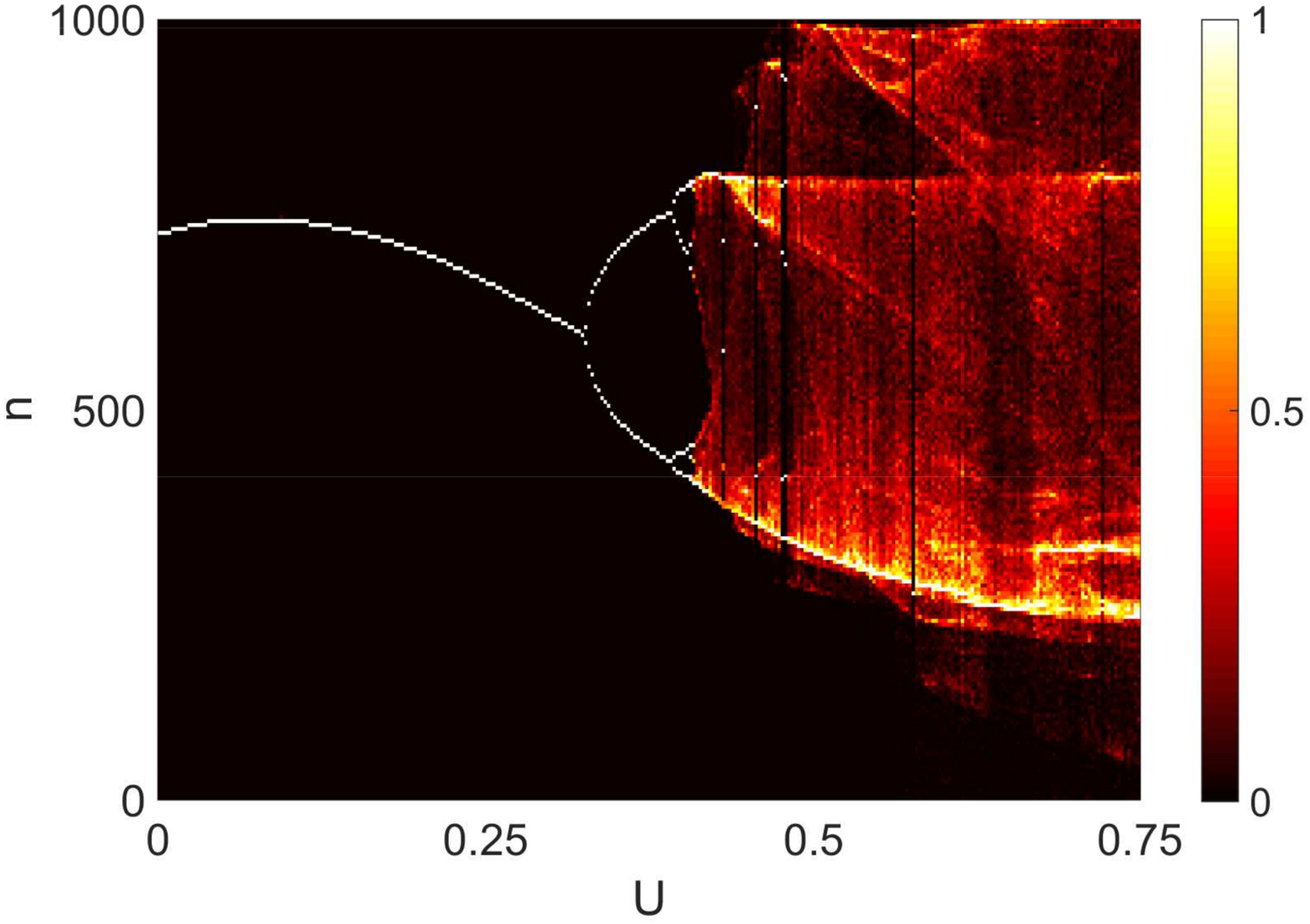}%
(b)  \includegraphics[angle=270,width=0.95\columnwidth]{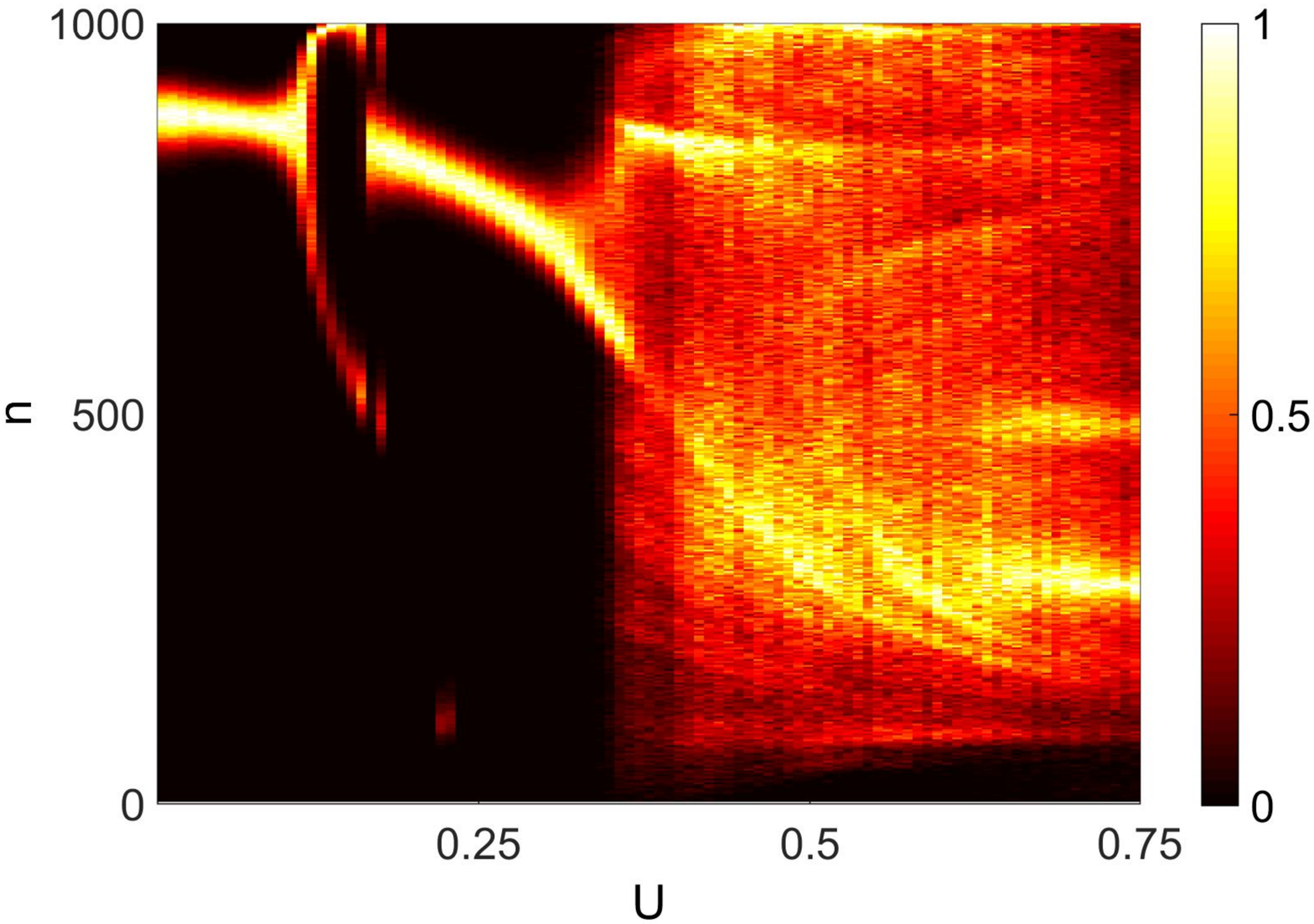}\\%

	\caption{\label{fig:4} 
    One-parameter bifurcation diagrams for the (a) classical and (b) quantum 
    periodically modulated systems, $J=-1, E=1, A=1.5, T=1, \gamma=0.1$. 
    In both cases, stroboscopic expectation values of the number of particles on the first site were recorded during $2000$ periods (after an equal transient time) and plotted as a color-coded histogram, with the maximal element normalized to one for each value of $U$. In the quantum case, the data were additionally collected from $8$ independent realizations.} 
 \end{figure*}

Next, we introduce a  bias, $E>0$, that favors transitions of particles to the second site. 
In classical dynamical systems, removing a symmetry of equations breaks a pitchfork bifurcation. 
Generically, in the super-critical case, a stable equilibrium avoids the bifurcation, while another pair of stable and unstable states are born in what becomes the saddle-node bifurcation. 
This picture is reproduced in the quantum case, see Fig.~\ref{fig:2}(b). 
Again, the new maximum closely follows the emerging stable equilibrium of the mean-field system (note that it also exists in the region  $U \in  [0.6, 0.7]$, 
but is not seen due to the limited color resolution). 

Returning to the symmetric case, $E=0$, 
we detect a richer bifurcation picture for weak hopping, $J\ll1$. 
There it is possible to observe even three maxima in the quantum bifurcation diagram, 
that emerge after two simultaneous saddle-node bifurcations as the interaction increases, Fig.~\ref{fig:2}(c). 
The maximum in the middle and its classical stable equilibrium counterpart then disappear in the sub-critical pitchfork bifurcation. Note, however, that there is a substantial mismatch in the position of this bifurcation on the $U$ axis. This indicates that the finite $N$ quantum system is not exactly described by the mean-field model. 

The drop in the magnitude of the second largest real part eigenvalue and purity of the stationary solution after quantum bifurcations is reproduced in all cases; it  appears to be a generic feature, Fig.\ref{fig:2}(e,f).

We underline that the observed quantum bifurcations are all manifested through the structural changes in the stationary density matrix rather than the loss of dynamical stability of this quantum steady state or its uniqueness. Therefore, the conditions of bifurcations in quantum systems must be different from those in classical dynamical systems. All the three considered types of bifurcations involve changes in the number of maxima along the diagonal of the stationary density matrix. Thus, a suitable bifurcation condition is either (i) the coincidence of the three consecutive diagonal 
elements, $\varrho_{n-1,n-1}=\varrho_{n,n}=\varrho_{n+1,n+1}$ or (ii)  disappearance of the first and second order finite differences, $\partial_n\varrho=\partial_{nn}\varrho=0$, 
while $\varrho_{n,n} \neq 0$.

Using  these definitions, we produce two-parameter quantum bifurcation diagrams, on $(U,E)$ and $(U,J)$ plains, and compare to those of the classical system, Fig.~\ref{fig:3}. The obtained results demonstrate an excellent correspondence, safe for the case of weak hopping, $J\ll1$, where noticeable mismatch has already been noticed.

\section{Chaotic dynamics}

Periodic driving enriches dynamics of an open quantum system and provides  possibility to create chaotic regimes in the corresponding mean-field system. We numerically integrated  Eq.(\ref{eq:5}) and recorded the mean number of particles in the first site stroboscopically, $n(m T)=N \big[1+\cos\vartheta(m T)\big]/2$, $m\in\mathbb{N}^+$, after some transient time. For the set of parameters $J=-1, E=1, A=1.5, T=1, \gamma=0.1, N=10^3$, the period-doubling route to chaos and the further development of the chaotic attractor is observed, see Fig.~\ref{fig:4}(a).

To resolve the fine structure of quantum 'chaotic attractors', we needed  a considerably greater number of particles  than previously, in the stationary case. 
Due to the growing dimension $N+1$ of the system Hilbert space, the direct numerical integration of Eq. (\ref{eq:1}) or its Floquet analysis becomes unfeasible, and we resort to the quantum trajectory method  \cite{zoller,dali,plenio}.   We implemented this method to calculate  expectation values of the number of particles on the first site at the stroboscopic instances of time. By using the sampled data points, we produce a histogram that corresponds to the diagonal elements of the asymptotic density matrix. Similar to the stationary bifurcations, the development of the quantum chaotic attractor goes through the emergence of new maxima 
in the diagonal matrix elements, Fig.~\ref{fig:4}(b). Remarkably, the structure of the chaotic attractor of the mean-field model are also reproduced. 

Finally, we also observe a quantum bifurcation at $U \sim 0.15$ that does not have a classical counterpart. Previously, we have noticed some mismatches in the bifurcation curves in the stationary case (Fig.~\ref{fig:3}), which is natural, since there is no exact identity between the finite particle quantum system and its mean-field approximation. Nevertheless, this time we face an example of a pure quantum bifurcation that persists up to surprisingly high particle numbers while being absent in the classical case.

\section{Conclusions}

By using a scalable many-body model, we have shown that the asymptotic density matrix of an open quantum system can be used to plot a quantum bifurcation diagram.
Our approach remains  valid after unraveling the original deterministic Markovian evolution of the density matrix into a set of quantum trajectories. The diagram is then calculated by performing the standard Monte-Carlo sampling.

The proposed  approach is also technically efficient because it allows one for skipping calculations of quasi-classical phase space distributions, Husimi or Wigner-like \cite{stock}.
The latter are numerically time-consuming operations; moreover, for the used model, calculation of Husimi distribution is not feasible when $N > 10^3$. At the same time, the quantum trajectory 
method allows to calculated bifurcation diagram for the system with several thousands of bosons.

There is an intriguing observation; namely, we find a quantum bifurcation which is absent in the mean-filed model and remains robust under increase of the boson number $N$. Here we can only speculate that this transition has a pure quantum nature and can not be explained in classical terms. However, this is an issue for further studies.

\section{Acknowledgments}\label{acknowledgment}

The authors acknowledge support of the Russian Science Foundation grant No. 15-12-20029. Numerical experiments were carried out at the Lobachevsky University Supercomputer.


\end{document}